\documentclass[12pt, draftclsnofoot, onecolumn]{IEEEtran}
\usepackage{lineno}
\usepackage{cite}
\usepackage{hyperref}
\usepackage{amsmath,amssymb,amsfonts}
\usepackage{amsmath}
\usepackage{amsthm}
\DeclareMathOperator*{\argmax}{argmax}

\usepackage{graphicx}
\usepackage{textcomp}
\usepackage{xcolor}
\usepackage{graphicx}
\usepackage{float}
\usepackage{subfigure}
\usepackage{amsmath}
\usepackage{amsfonts,amssymb}
\usepackage{mathrsfs}
\usepackage{mathtools}
\usepackage{multirow}
\usepackage{algpseudocode}
\makeatletter
\newif\if@restonecol
\makeatother

\usepackage[linesnumbered,ruled,vlined]{algorithm2e}
\usepackage{algpseudocode}

\usepackage{setspace}
\usepackage{footmisc}
\usepackage[justification=centering]{caption}

\usepackage{mathtools}
\usepackage{dsfont}
\usepackage{bbm}
\newtheorem{remark}{Remark}
\theoremstyle{definition}
\newtheorem{corollary}{Corollary}

\newtheorem{theorem}{Theorem}

\newtheorem{lemma}{Lemma}

\makeatletter
\newcommand{\biggg}{\bBigg@{3}}
\newcommand{\Biggg}{\bBigg@{3.5}}
\makeatother
\usepackage{bm}
\makeatother
\begin{document}

\title{On the Ergodic Mutual Information of Keyhole MIMO Channels With Finite-Alphabet Inputs}
\author{Chongjun~Ouyang, Ali Bereyhi, Saba Asaad, Ralf~R.~M\"{u}ller, Julian~Cheng, and Hongwen~Yang
\thanks{C. Ouyang and H. Yang are with the School of Information and Communication Engineering, Beijing University of Posts and Telecommunications, Beijing, 100876, China (e-mail: \{DragonAim,yanghong\}@bupt.edu.cn).}
\thanks{A. Bereyhi, S. Asaad, and R. R. M\"{u}ller are with the Institute for Digital Communications, Friedrich-Alexander-Universit\"{a}t Erlangen-N\"{u}rnberg, 91058, Erlangen, Germany (e-mail: \{ali.bereyhi,saba.asaad,ralf.r.mueller\}@fau.de).}
\thanks{J. Cheng is with the School of Engineering, The University of British Columbia, Kelowna, BC V1V 1V7, Canada (email: julian.cheng@ubc.ca).}}
\maketitle

\begin{abstract}
This letter studies the ergodic mutual information (EMI) of keyhole multiple-input multiple-output channels having finite-alphabet input signals. The EMI is first investigated for single-stream transmission considering both cases with and without the channel state information at the transmitter. Then, the derived results are extended to the scenario of multi-stream transmission. Asymptotic analyses are performed in the regime of high signal-to-noise ratio (SNR). The high-SNR EMI is shown to converge to a constant with its rate of convergence determined by the diversity order. On this basis, the influence of the keyhole effect on the EMI is discussed. The analytical results are validated by numerical simulations.
\end{abstract}


\section{Introduction}
\begin{figure}[!t]
\centering
\setlength{\abovecaptionskip}{0pt}
\includegraphics[height=0.3\textwidth]{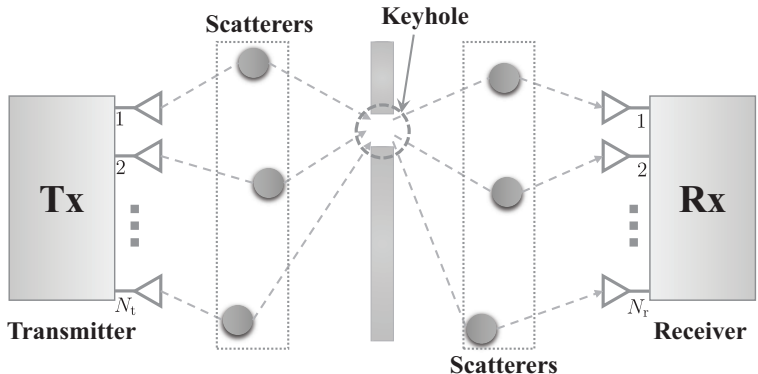}
\caption{Illustration of a keyhole MIMO channel}
\label{System_Model}
\vspace{-20pt}
\end{figure}

Multiple-input multiple-output (MIMO) systems are known to boost the spectral efficiency (SE) of wireless channels in comparison to conventional single-antenna systems. Yet, practical MIMO systems may suffer from severe degradation of the SE, due to channel degeneration. One of such phenomena is termed the keyhole effect which may arise in a hallway or tunnel with the electromagnetic waves propagating through the same hole as shown in {\figurename} {\ref{System_Model}}; see \cite{Chizhik2002,Ngo2017,Speidel2007,Maaref2008,Almers2006,Akemann2013} and the references therein. This effect is observed in various applications; for instance, in vehicle-to-vehicle communications under dense urban environments \cite{Zhang2022}. The existence of this effect was initially predicted in theory \cite{Chizhik2002} and then validated by empirical measurements \cite{Almers2006}. In contrast to traditional MIMO channels, keyhole channels generally characterize rank-deficient MIMO channels, which may have sufficient scattering around the transceivers, but due to other propagation effects, such as diffraction, the channel matrix might exhibit only low rank.

Theoretically, the keyhole effect can remove the spatial multiplexing gain of MIMO channels \cite{Chizhik2002}. It hence models the worst-case propagation environment for MIMO systems from the SE perspective. In general, the system SE is proportional to the achievable input-output mutual information (MI) of the channel \cite{Lozano2018}. Consequently, analyzing the MI of keyhole MIMO channels can benchmark the worst-case SE of multiple-antenna systems. Motivated by this, several studies analyzed the MI of keyhole MIMO channels for Gaussian distributed input signals \cite{Maaref2008,Almers2006,Speidel2007,Akemann2013,Zhang2022,Ngo2017}. Particularly, the MI achieved by Gaussian inputs was analyzed in single-user ergodic case \cite{Maaref2008,Almers2006,Speidel2007,Akemann2013}, multi-user ergodic case \cite{Ngo2017}, and the single-user outage case \cite{Zhang2022}. Yet, practical transmit signals are often taken from finite constellation alphabets, e.g., quadrature amplitude modulation (QAM). These finite constellations yield reduced MI, especially in the high signal-to-noise ratio (SNR) regime \cite{Lozano2018,Rodrigues2010,Ouyang2020,Alvarado2014}. Despite its importance, analysis of the ergodic MI (EMI) for keyhole MIMO channels with finite input constellations has been left open.

This letter studies the EMI of keyhole MIMO channels with finite-alphabet inputs under Nakagami-$m$ fading. The main contributions of this work are as follows: 1) We derive novel expressions of the EMI under single-stream transmission (SST) by considering perfect CSI at the receiver and both cases with and without the CSI at the transmitter (CSIT); 2) We extend the scenario of SST to the scenario of multi-stream transmission (MST) and study the EMI under three typical precoding schemes; 3) We characterize the EMI in the high-SNR region and determine the diversity order of the system, which enables us to estimate the influence of the keyhole effect\footnote{We comment that also for MIMO channels without keyholes, there has been very limited work on characterizing the high-SNR asymptotic behaviours of the EMI achieved by finite-alphabet inputs. Yet, this can be done by using the approach proposed in this work, which will be considered in the future.}. Compared with our previous work \cite{Ouyang2020} that focused more on approximating the EMI in single-antenna systems and neglected the high-SNR analyses, this letter gains more insights into the influence of finite-alphabet on the EMI in MIMO keyhole channels.

\section{System Model}
Consider the point-to-point keyhole MIMO channel illustrated in {\figurename} {\ref{System_Model}}, where an $N_{\rm{t}}$-antenna transmitter (Tx) sends wireless signals to an $N_{\rm{r}}$-antenna receiver (Rx). The received signal is given by
{\setlength\abovedisplayskip{2pt}
\setlength\belowdisplayskip{2pt}
\begin{align}
{\mathbf{y}}=\sqrt{\bar\gamma}{\mathbf{H}}{\mathbf{s}}+{\mathbf{n}},
\end{align}
}where ${\mathbf{H}}\in{\mathbbmss{C}}^{N_{\rm{r}}\times N_{\rm{t}}}$ represents the channel matrix with $N_{\rm{t}}>1$ and $N_{\rm{r}}>1$, ${\mathbf{s}}\in{\mathbbmss{C}}^{N_{\rm{t}}\times1}$ denotes the transmit signal satisfying ${\mathbbmss{E}}\left\{{\mathbf{s}}^{\mathsf H}{\mathbf{s}}\right\}=1$, $\bar\gamma$ denotes the transmit SNR, and ${\mathbf{n}}\sim{\mathcal{CN}}\left({\mathbf{0}},{\mathbf{I}}_{N_{\rm{r}}}\right)$ is additive white Gaussian noise (AWGN).

Considering the spatial structure of keyhole MIMO channels, we have ${\mathbf{H}}={\mathbf{h}}_{\rm{r}}{\mathbf{h}}_{\rm{t}}^{\mathsf H}$ for ${\mathbf{h}}_{\rm{r}}\in{\mathbbmss{C}}^{N_{\rm{r}}\times 1}$ and ${\mathbf{h}}_{\rm{t}}\in{\mathbbmss{C}}^{N_{\rm{t}}\times 1}$, where
{\setlength\abovedisplayskip{2pt}
\setlength\belowdisplayskip{2pt}
\begin{align}
&{\mathbf{h}}_{\rm{r}}=\left[\sqrt{\alpha_1}{\rm{e}}^{{\rm{j}}\phi_1},\ldots,\sqrt{\alpha_{N_{\rm{r}}}}{\rm{e}}^{{\rm{j}}\phi_{N_{\rm{r}}}}\right]^{\mathsf T}\in{\mathbbmss{C}}^{N_{\rm{r}}\times 1},\\
&{\mathbf{h}}_{\rm{t}}=\left[\sqrt{\beta_1}{\rm{e}}^{{\rm{j}}\psi_1},\ldots,\sqrt{\beta_{N_{\rm{t}}}}{\rm{e}}^{{\rm{j}}\psi_{N_{\rm{t}}}}\right]^{\mathsf T}\in{\mathbbmss{C}}^{N_{\rm{t}}\times 1},
\end{align}
}denote the keyhole-to-Rx and keyhole-to-Tx channel vectors, respectively, which are statistically independent of each other \cite{Chizhik2002}\footnote{It is worth mentioning that the keyhole channel is also influenced by the size of the keyhole. Intuitively, the keyhole effect is more pronounced when the keyhole's physical size approximately equals or is even smaller than the wavelength \cite{Almers2006}. Unfortunately, a quantitative characterization of the influence of the keyhole's size on the channel is still open.}. We assume that all entries in the vector ${\mathbf{h}}_{\rm{r}}$ are independent and identically distributed (i.i.d.), i.e., the phases $\phi_a$ for $a\in\{ 1, \ldots, N_{\rm{r}}\}$ are uniformly distributed on $\left[0,2\pi\right)$ and the magnitudes $\sqrt{\alpha_a}$ follow the Nakagami-$m$ distribution with the probability density function (PDF) of $\alpha_a$ given by $f\left(x;m_{\rm{r}},m_{\rm{r}}\right)$. Here,
{\setlength\abovedisplayskip{2pt}
\setlength\belowdisplayskip{2pt}
\begin{align}
f\left(x;c,d\right)\triangleq\frac{1}{\Gamma\left(c\right)}x^{c-1}{\rm{e}}^{-dx}d^c,x\geq0,
\end{align}
}where $\Gamma\left(x\right)\triangleq\int_{0}^{\infty}t^{x-1}{\rm{e}}^{-t}{\rm{d}}t$ is the gamma function \cite{Paris2010}, and $m_{\rm{r}}\geq\frac{1}{2}$ indicates the fading severity. Likewise, we assume that the keyhole-to-Tx channel undergoes i.i.d. Nakagami-$m$ fading; thus, the PDF of the magnitudes $\beta_b$ for $b\in\{ 1, \ldots, N_{\rm{t}}\}$ is given by $f\left(x;m_{\rm{t}},m_{\rm{t}}\right)$ for some fading severity $m_{\rm{t}}\geq\frac{1}{2}$ and the phases $\psi_b$ are uniformly distributed on $\left[0,2\pi\right)$. It is worth noting that the Nakagami-$m$ model is a generalization of the statistical model used in \cite{Almers2006,Zhang2022,Ngo2017}, which has been illustrated to fit better with empirical data.

\section{Single-Stream Transmission}
We start the analysis by considering the SST. The transmitted signal is given by ${\mathbf{s}}={\mathbf{w}}x$, where ${\mathbf{w}}\in{\mathbbmss{C}}^{N_{\rm{r}}\times1}$ denotes the precoding vector satisfying $\left\|{\mathbf{w}}\right\|^2=1$ and $x\in{\mathbbmss{C}}$ is the transmitted symbol. We assume that $x$ satisfies the power constraint ${\mathbbmss{E}}\{\left|x\right|^2\}=1$ and is taken from a finite constellation alphabet $\mathcal X$ consisting of $M$ points, i.e., ${\mathcal{X}}=\left\{\mathsf{x}_g\right\}_{g=1}^{M}$. The $g$th symbol in $\mathcal{X}$, i.e, $\mathsf{x}_g$, is transmitted with probability $p_g$, $0 < p_g < 1$, and the vector of probabilities ${\mathbf{p}}_{\mathcal{X}}\triangleq[p_1,\cdots,p_M]\in{\mathbbmss{C}}^{1\times M}$ is called the input distribution with $\sum_{g=1}^{M}p_g=1$.

The derivation of EMI for the SST (SST-EMI) in a fading keyhole MIMO channel is best understood by specifying the MI of a scalar Gaussian channel with finite-alphabet inputs. To this end, consider the scalar AWGN channel
{\setlength\abovedisplayskip{2pt}
\setlength\belowdisplayskip{2pt}
\begin{equation}\label{AWGN_Channel}
Y=\sqrt{\gamma}X+Z,
\end{equation}
}where $Z\sim{\mathcal {CN}}\left(0,1\right)$ is AWGN, $X$ is the channel input taken from the alphabet $\mathcal X$ subject to the input distribution ${\mathbf{p}}_{\mathcal{X}}$, and $\gamma$ is the SNR. For this channel, the MI is given by \cite{Lozano2018}
{\setlength\abovedisplayskip{2pt}
\setlength\belowdisplayskip{2pt}
\begin{equation}
\begin{split}
I_{M}^{\mathcal X}\left(\gamma\right)&=H_{{\mathbf{p}}_{\mathcal{X}}}-\frac{1}{\pi}\sum\nolimits_{g=1}^{M}\int_{\mathbbmss C}p_g{\rm e}^{-\left|u-\sqrt{\gamma}{\mathsf{x}}_g\right|^2}\\
&\times\log_2{\left(\sum\nolimits_{{g'}=1}^{M}\frac{p_{g'}}{p_g}{\rm e}^{\left|u-\sqrt{\gamma}{\mathsf{x}}_g\right|^2-\left|u-\sqrt{\gamma}{\mathsf{x}}_{g'}\right|^2}\right)}{\rm d}u,
\end{split}
\end{equation}
}where $H_{{\mathbf{p}}_{\mathcal{X}}}$ is the entropy of the input distribution ${\mathbf{p}}_{\mathcal{X}}$ in bits. By a straightforward extension of this result to a single-input vectorized channel, it is shown that the SST-EMI achieved by maximum ratio combining is given by
{\setlength\abovedisplayskip{2pt}
\setlength\belowdisplayskip{2pt}
\begin{align}
{\mathcal{I}}_{M}^{\mathcal{X}}={\mathbbmss{E}}\{I_{M}^{\mathcal X}({\bar\gamma}\left\|{\mathbf{h}}_{\rm{r}}\right\|^2\left|{\mathbf{h}}_{\rm{t}}^{\mathsf H}{\mathbf{w}}\right|^2)\}.
\end{align}
}It is worth noting that the EMI is a function of the precoding vector ${\mathbf w}$. In the sequel, we will analyze the SST-EMI based on the availability of CSIT.

\subsection{SST Without CSIT}
With no CSIT, the transmitter applies uniform beamforming, i.e., ${\mathbf{w}}=\frac{1}{\sqrt{N_{\rm{t}}}}{\mathbf{1}}$, where ${\mathbf{1}}\triangleq\left[1,\cdots,1\right]^{\mathsf{T}}$. In this case, we have ${\mathcal{I}}_{M}^{\mathcal{X}}
={\mathbbmss{E}}\left\{I_{M}^{\mathcal X}\left(S_1{\bar\gamma}/{N_{\rm{t}}}\right)\right\}$, where $S_1=\left\|{\mathbf{h}}_{\rm{r}}\right\|^2\left|{\mathbf{h}}_{\rm{t}}^{\mathsf H}{\mathbf{1}}\right|^2$. To characterize the EMI, we follow three major steps which are illustrated in the sequel.
\subsubsection{Channel Statistics}
At the first step, we derive the PDF of $S_1$. The statistical independence of ${\mathbf{h}}_{\rm{t}}$ and ${\mathbf{h}}_{\rm{r}}$ concludes that $A=\left\|{\mathbf{h}}_{\rm{r}}\right\|^2$ and $B=\left|{\mathbf{h}}_{\rm{t}}^{\mathsf H}{\mathbf{1}}\right|^2$ are mutually independent. It follows that the PDF of the product $S_1=AB$ can be calculated as $f_{S_1}\left(x\right)=\int_{0}^{\infty}f_{B}\left(\frac{x}{y}\right)f_A\left(y\right)\frac{1}{y}{\rm{d}}y$, where $f_{A}(\cdot)$ and $f_{B}(\cdot)$ denote the PDFs of $A$ and $B$, respectively. Yet, due to the intractability of $\left|{\mathbf{h}}_{\rm{t}}^{\mathsf H}{\mathbf{1}}\right|^2$, a closed-form expression for its PDF is only available when $m_{\rm{t}}$ is an integer \cite{Karagiannidis2006}. Accordingly, we let $m_{\rm{t}}$ be an integer in order to facilitate the subsequent analyses. The following two lemmas are then employed to characterize $S_1$.
\vspace{-5pt}
\begin{lemma}\label{Lemma_NCSIT_SS_Nakagami}
Define an operator ${\mathcal{F}}\left\langle{\cdot}\right\rangle$ as
{\setlength\abovedisplayskip{2pt}
\setlength\belowdisplayskip{2pt}
\begin{equation}
{\mathcal{F}}\left\langle{Q}\right\rangle\triangleq\sum_{i_1=0}^{m_{\rm{t}}-1}\cdots\sum_{i_{N_{\rm{t}}}=0}^{m_{\rm{t}}-1}\sum_{h=0}^{S_{N_{\rm{t}}}}
\frac{\left(-S_{N_{\rm{t}}}\right)_hS_{N_{\rm{t}}}!
Y_{N_{\rm{t}}}{Q}}{X_{N_{\rm{t}}}\left(h!\right)^2U_{N_{\rm{t}}}^{S_{N_{\rm{t}}}}},
\end{equation}
}where $X_{N_{\rm{t}}}=\prod_{k=1}^{N_{\rm{t}}}\left(\frac{\left(i_k!\right)^2}{\left(1-m_{\rm{t}}\right)_{i_k}}\right)$, $S_{N_{\rm{t}}}=\sum_{k=1}^{N_{\rm{t}}}i_k$, $Y_{N_{\rm{t}}}=\prod_{k=1}^{N_{\rm{t}}}\left(\frac{1}{4m_{\rm{t}}}\right)^{i_k}$, $U_{N_{\rm{t}}}=\sum_{k=1}^{N_{\rm{t}}}\frac{1}{4m_{\rm{t}}}$, and $\left(z\right)_n\triangleq\frac{\Gamma\left(z+n\right)}{\Gamma\left(z\right)}$ is the Pochhammer symbol \cite[Eq. (5.2.5)]{Paris2010} with $\left(-z\right)_n=\left(-1\right)^n\left(z-n+1\right)_n$. Then, the PDF of $B=\left|{\mathbf{h}}_{\rm{t}}^{\mathsf H}{\mathbf{1}}\right|^2$ can be written as $f_{B}\left(x\right)={\mathcal{F}}\left\langle{{\rm{e}}^{-\frac{x}{4U_{N_{\rm{t}}}}}x^h\left(4U_{N_{\rm{t}}}\right)^{-h-1}}\right\rangle$.
\end{lemma}
\vspace{-5pt}
\begin{IEEEproof}
Please refer to \cite{Karagiannidis2006} for more details.
\end{IEEEproof}
\vspace{-5pt}
\begin{lemma}\label{Lemma_NCSIT_SS_PDF}
The PDF of $S_1$ is given by
{\setlength\abovedisplayskip{2pt}
\setlength\belowdisplayskip{2pt}
\begin{equation}\label{NCSIT_SS_PDF}
\begin{split}
f_{S_1}\left(x\right)&={\mathcal{F}}\left\langle
\frac{2}{\Gamma\left(N_{\rm{r}}m_{\rm{r}}\right)}\left({m_{\rm{r}}x}/{\left(4U_{N_{\rm{t}}}\right)}\right)^{\frac{N_{\rm{r}}m_{\rm{r}}+h+1}{2}}\right.\\
&\times \left. x^{-1}K_{N_{\rm{r}}m_{\rm{r}}-h-1}\left(2\sqrt{{m_{\rm{r}}x}/{\left(4U_{N_{\rm{t}}}\right)}}\right)\right\rangle,
\end{split}
\end{equation}
}where $K_{\nu}\left(\cdot\right)$ is the $\nu$th order modified Bessel function of the second kind \cite[Eq. (10.31.1)]{Paris2010}.
\end{lemma}
\vspace{-5pt}
\begin{IEEEproof}
Since $\left\{\sqrt{\alpha_a}\right\}_{a=1}^{N_{\rm{r}}}$ are $N_{\rm{r}}$ i.i.d. Nakagami-$m$ variables, the PDF of ${A}=\sum_{a=1}^{N_{\rm{r}}}\alpha_a$ can be written as $f_A\left(x\right)=f\left(x;N_{\rm{r}}m_{\rm{r}},m_{\rm{r}}\right)$. Aided with the integral identity in \cite[Eq. (10.32.10)]{Paris2010}, we finally conclude the desired PDF in \eqref{NCSIT_SS_PDF}.
\end{IEEEproof}

\subsubsection{Explicit Analysis}
In the second step, we invoke Lemma \ref{Lemma_NCSIT_SS_PDF} to derive an approximation for the SST-EMI.
\vspace{-5pt}
\begin{theorem}\label{Theorem_NCSIT_SS_EMI_Appr}
For SST-EMI achieved without CSIT, the following approximation becomes exact as the complexity-vs-accuracy tradeoff parameter $V$ approaches infinity:
{\setlength\abovedisplayskip{2pt}
\setlength\belowdisplayskip{2pt}
\begin{align}
{\mathcal{I}}_{M}^{\mathcal{X}}\approx{\mathcal{F}}\left\langle\sum_{k=1}^{V}\sum_{l=1}^{V}
\frac{w_kw_lI_M^{\mathcal{X}}\left(\frac{4U_{N_{\rm{t}}}\bar\gamma t_kt_l}{m_{\rm{r}}N_{\rm{t}}}\right)}{\Gamma\left(N_{\rm{r}}m_{\rm{r}}\right)t_l^{-j}t_k^{1-N_{\rm{r}}m_{\rm{r}}}}
\right\rangle,\label{NCSIT_SS_EMI_Appr}
\end{align}
}where $\left\{w_i\right\}$ and $\left\{t_i\right\}$ denote the weight and abscissa factors of Gauss–Laguerre integration.
\end{theorem}
\vspace{-5pt}
\begin{IEEEproof}
The EMI can be calculated as
{\setlength\abovedisplayskip{2pt}
\setlength\belowdisplayskip{2pt}
\begin{align}\label{EMI_SS_Exact_NCSIT_Exact_Calculation}
{\mathcal{I}}_{M}^{\mathcal{X}}=\int_{0}^{\infty}\left(\int_{0}^{\infty}f_{B}\left(\frac{x}{y}\right)\frac{f_{A}\left(y\right)}{y}{\rm{d}}y\right)I_{M}^{\mathcal X}\left(\frac{x\bar\gamma}{N_{\rm{t}}} \right){\rm d}x.
\end{align}
}We use the Gauss–Laguerre quadrature method \cite[Eq. (3.5.27)]{Paris2010} to calculate the two integrals in \eqref{EMI_SS_Exact_NCSIT_Exact_Calculation} successively. This leads to the approximate expression shown in \eqref{NCSIT_SS_EMI_Appr}.
\end{IEEEproof}
Note that given a target approximation precision, quantifying the relationship between the required value of $V$ and other system parameters, such as $M$ and $\bar\gamma$, is challenging. By numerical simulation, we find out that setting $V=200$ can generally achieve an approximation precision of $10^{-14}$.
\subsubsection{Asymptotic Analysis}
In the last step, we investigate the asymptotic behaviour of the EMI. It is worth noting that the MIMO keyhole channel does not always harden under the asymptotic condition when $N_{\rm{t}}~{\text{or}}~N_{\rm{r}}\rightarrow\infty$ \cite{Ngo2017}. This makes it challenging to gain further insights into the EMI by setting $N_{\rm{t}}~{\text{or}}~N_{\rm{r}}\rightarrow\infty$. As a compromise, more attention will be paid to the asymptotic limit in which the SNR approaches infinity, i.e., $\bar\gamma\rightarrow\infty$. The result is given in Theorem \ref{Theorem_EMI_Asy_SS_NCSIT_Asym_Basic}.
\vspace{-5pt}
\begin{theorem}\label{Theorem_EMI_Asy_SS_NCSIT_Asym_Basic}
   Let $N_{\rm{r}}m_{\rm{r}}\neq h+1$ for $h\in\left\{0,\cdots,N_{\rm{t}}(m_{\rm{t}}-1)\right\}$. When $\bar\gamma\rightarrow\infty$, the EMI achieved without CSIT can be characterized as ${\mathcal{I}}_{M}^{\mathcal{X}}\simeq{H_{{\mathbf{p}}_{\mathcal{X}}}}-\left({\mathcal{G}}_{\rm{a}}{\bar\gamma}\right)^{-{\mathcal{G}}_{\rm{d}}}$, where ${\mathcal{G}}_{\rm{d}}=1$ and
{\setlength\abovedisplayskip{2pt}
\setlength\belowdisplayskip{2pt}
\begin{align}
{\mathcal{G}}_{\rm{a}}^{-1}\!=\!\sum\limits_{i_1=0}^{m_{\rm{t}}-1}\!\!\cdots\!\!\sum\limits_{i_{N_{\rm{t}}}=0}^{m_{\rm{t}}-1}
\!\frac{U_{N_{\rm{t}}}^{-S_{N_{\rm{t}}}}S_{N_{\rm{t}}}!
Y_{N_{\rm{t}}}\hat{\mathcal{M}}\left(2\right)m_{\rm{t}}m_{\rm{r}}\log_2{\rm{e}}}{\left(N_{\rm{r}}m_{\rm{r}}-1\right)\prod_{k=1}^{N_{\rm{t}}}\left(\frac{\left(i_k!\right)^2}
{\left(1-m_{\rm{t}}\right)_{i_k}}\right)}.\label{Ga}
\end{align}
}Here, $\hat{\mathcal{M}}\left({x}\right)\triangleq{\mathcal M}\left[{\mathrm{mmse}}_{M}^{\mathcal X}\left(t\right);{x}\right]$,
where ${\mathrm{mmse}}_{M}^{\mathcal X}\left(t\right)$ denotes the minimum mean square error (MMSE) in estimating $X$ in \eqref{AWGN_Channel} from $Y$. Moreover, ${\mathcal M}\left[p\left(t\right);z\right]\triangleq\int_{0}^{\infty}t^{z-1}p\left(t\right){\rm d}t$ denotes the Mellin transform of $p\left(t\right)$ \cite{Flajolet1995}.
\end{theorem}
\vspace{-5pt}
\begin{IEEEproof}
The proof is given in Appendix \ref{Proof_Theorem_EMI_Asy_SS_NCSIT_Asym_Basic}.
\end{IEEEproof}
\vspace{-5pt}
\begin{remark}\label{remark1}
The results in Theorem \ref{Theorem_EMI_Asy_SS_NCSIT_Asym_Basic} suggest that the EMI achieved by finite-alphabet input signals converges to ${H_{{\mathbf{p}}_{\mathcal{X}}}}$ as the SNR increases and its rate of convergence (ROC) is determined by the diversity order ${\mathcal{G}}_{\rm{d}}$ and the array gain ${\mathcal{G}}_{\rm{a}}$.
\end{remark}
\vspace{-5pt}

\subsection{SST With CSIT}
With CSIT, we can apply maximal ratio transmission (MRT) at the transmitter, i.e., ${\mathbf{w}}=\frac{1}{\left\|{\mathbf{h}}_{\rm{t}}\right\|}{\mathbf{h}}_{\rm{t}}$. Hence, the EMI is given by ${\mathcal{I}}_{M}^{\mathcal{X}}
={\mathbbmss{E}}\left\{I_{M}^{\mathcal X}\left({\bar\gamma}S_2\right)\right\}$, where $S_2=\left\|{\mathbf{h}}_{\rm{r}}\right\|^2\left\|{\mathbf{h}}_{\rm{t}}\right\|^2$. Similar to the previous case, we characterize the SST-EMI in three steps.
\subsubsection{Channel Statistics}
Using similar steps as those outlined in the proof of Lemma \ref{Lemma_NCSIT_SS_PDF}, we arrive at the following lemma.
\vspace{-5pt}
\begin{lemma}\label{Lemma_CSIT_SS_PDF}
The PDF of $S_2$ is given by
{\setlength\abovedisplayskip{2pt}
\setlength\belowdisplayskip{2pt}
\begin{align}
f_{S_2}\left(x\right)\!=\!\frac{2\!\left(m_{\rm t}m_{\rm r}\right)^{\frac{N_{\rm t}m_{\rm t}+N_{\rm r}m_{\rm r}}{2}}\!K_{N_{\rm r}m_{\rm r}-N_{\rm t}m_{\rm t}}\left(2\sqrt{m_{\rm t}m_{\rm r}x}\right)}{\Gamma\left(N_{\rm t}m_{\rm t}\right)\Gamma\left(N_{\rm r}m_{\rm r}\right)x^{1-\frac{N_{\rm t}m_{\rm t}+N_{\rm r}m_{\rm r}}{2}}}.\nonumber
\end{align}
}\end{lemma}
\vspace{-5pt}
\begin{IEEEproof}
The proof is similar to the one given for Lemma \ref{Lemma_NCSIT_SS_PDF}. We hence omit it.
\end{IEEEproof}

\subsubsection{Explicit Analysis}
The EMI with CSIT is given by ${\mathcal{I}}_{M}^{\mathcal{X}}=\int_{0}^{\infty}f_{S_2}\left(x\right)I_{M}^{\mathcal X}\left({\bar\gamma}x\right){\rm{d}}x$. By following the same steps as those taken in the proof of Theorem \ref{Theorem_NCSIT_SS_EMI_Appr}, we conclude the following approximation for ${\mathcal{I}}_{M}^{\mathcal{X}}$:
{\setlength\abovedisplayskip{2pt}
\setlength\belowdisplayskip{2pt}
\begin{align}
{\mathcal{I}}_{M}^{\mathcal{X}}\approx\sum_{k=1}^{V}\sum_{l=1}^{V}\frac{w_kw_lt_j^{N_{\rm{r}}m_{\rm{r}}-1}t_l^{N_{\rm{t}}m_{\rm{t}}-1}}{\Gamma\left(N_{\rm{r}}m_{\rm{r}}\right)\Gamma\left(N_{\rm{t}}m_{\rm{t}}\right)}
I_M^{\mathcal{X}}\left(\frac{\bar\gamma t_kt_l}{m_{\rm{r}}m_{\rm{t}}}\right).\label{CSIT_SS_EMI_Appr}
\end{align}
}Similar to \eqref{NCSIT_SS_EMI_Appr}, the approximation in \eqref{CSIT_SS_EMI_Appr} becomes exact as the complexity-vs-accuracy tradeoff parameter $V$ tends infinity.

\subsubsection{Asymptotic Analysis}\label{section3c}
The limiting EMI with CSIT for asymptotically high SNRs is characterized as follows.
\vspace{-5pt}
\begin{theorem}\label{Theorem_EMI_Asy_SS_CSIT_Asym_Basic}
  Let $N_{\rm r}m_{\rm r}\neq N_{\rm t}m_{\rm t}$. When $\bar\gamma\rightarrow\infty$, the asymptotic EMI with CSIT satisfies ${\mathcal{I}}_{M}^{\mathcal{X}}\simeq{H_{{\mathbf{p}}_{\mathcal{X}}}}-\left({\mathcal{G}}_{\rm{a}}{\bar\gamma}\right)^{-{\mathcal{G}}_{\rm{d}}}$,
where ${\mathcal{G}}_{\rm{d}}=\min\left\{N_{\rm{t}}m_{\rm{t}},N_{\rm{r}}m_{\rm{r}}\right\}$ and
{\setlength\abovedisplayskip{2pt}
\setlength\belowdisplayskip{2pt}
\begin{align}
{\mathcal{G}}_{\rm{a}}=\frac{1}{m_{\rm{r}}m_{\rm{t}}}\left(\frac{\Gamma\left(N_{\rm{t}}m_{\rm{t}}\right)\Gamma\left(N_{\rm{r}}m_{\rm{r}}\right){\mathcal{G}}_{\rm{d}}\ln{2}}
{\Gamma\left(\left|N_{\rm{t}}m_{\rm{t}}-N_{\rm{r}}m_{\rm{r}}\right|\right)\hat{\mathcal{M}}\left({\mathcal{G}}_{\rm{d}}+1\right)}\right)^{1/{{\mathcal{G}}_{\rm{d}}}}.\nonumber
\end{align}
}\end{theorem}
\vspace{-5pt}
\begin{IEEEproof}
The proof is given by directly applying the method detailed in Appendix \ref{Proof_Theorem_EMI_Asy_SS_NCSIT_Asym_Basic}. We hence skip the details.
\end{IEEEproof}
\vspace{-5pt}
\begin{remark}\label{Compare_CSI}
The above result suggest that the diversity order in this case is a function of $\left\{N_{\rm{t}},N_{\rm{r}},m_{\rm{t}},m_{\rm{r}}\right\}$. By increasing the number of antennas, this expression can become larger than the one derived for the case without CSIT.
\end{remark}
\vspace{-5pt}
\subsection{Discussions on Keyhole Rank-Deficiency}
Consider a special case, in which the amplitudes of the channel coefficients follow the Rayleigh distribution, namely $m_{\rm{t}}=m_{\rm{r}}=1$. The MIMO channel matrix in this case has full rank, if there exist no keyholes \cite{Chizhik2002}. Using the method presented in Appendix \ref{Proof_Theorem_EMI_Asy_SS_NCSIT_Asym_Basic}, we can characterize the high-SNR SST-EMI in the keyhole and full-rank MIMO channels, respectively. Particularly, in the keyhole MIMO channel, the high-SNR SST-EMI achieved with and without CSIT can be written as ${\mathcal{I}}_{M,{\rm{c}},{\rm{r}}}^{\mathcal{X}}\simeq{H_{{\mathbf{p}}_{\mathcal{X}}}}-{\mathcal{O}}\left({\bar\gamma}^{-\min\left\{N_{\rm{t}},N_{\rm{r}}\right\}}\right)$\footnote{The notation $f(x)={\mathcal{O}}\left(g(x)\right)$ means that $\limsup_{x\rightarrow\infty}\frac{\left|f(x)\right|}{g(x)}<\infty$.} and ${\mathcal{I}}_{M,{\rm{n}},{\rm{r}}}^{\mathcal{X}}\simeq{H_{{\mathbf{p}}_{\mathcal{X}}}}-{\mathcal{O}}\left({\bar\gamma}^{-1}\right)$, respectively. Moreover, in full-rank MIMO channels, the high-SNR SST-EMI achieved with and without CSIT can be expressed as ${\mathcal{I}}_{M,{\rm{c}},{\rm{nk}}}^{\mathcal{X}}\simeq{H_{{\mathbf{p}}_{\mathcal{X}}}}-{\mathcal{O}}\left({\bar\gamma}^{-N_{\rm{r}}N_{\rm{t}}}\right)$ and ${\mathcal{I}}_{M,{\rm{n}},{\rm{nk}}}^{\mathcal{X}}\simeq{H_{{\mathbf{p}}_{\mathcal{X}}}}-{\mathcal{O}}\left({\bar\gamma}^{-N_{\rm{r}}}\right)$, respectively.
\vspace{-5pt}
\begin{remark}
Comparing ${\mathcal{I}}_{M,{\rm{c}},{\rm{r}}}^{\mathcal{X}}$ (or ${\mathcal{I}}_{M,{\rm{n}},{\rm{r}}}^{\mathcal{X}}$) with ${\mathcal{I}}_{M,{\rm{c}},{\rm{nk}}}^{\mathcal{X}}$ (or ${\mathcal{I}}_{M,{\rm{n}},{\rm{nk}}}^{\mathcal{X}}$), we conclude that the keyhole effect can reduce the diversity order of the SST-EMI.
\end{remark}
\vspace{-5pt}
We can extend the above results to a more generic case, where the rank of the channel matrix is smaller than $\min\left\{N_{\rm{r}},N_{\rm{t}}\right\}$, i.e., the channel matrix is rank-deficient. One example of such rank-deficient channels is a multi-keyhole MIMO channel whose number of keyholes is smaller than $\min\left\{N_{\rm{r}},N_{\rm{t}}\right\}$. Particularly, for fixed $N_{\rm{r}}$ and $N_{\rm{t}}$, the SST-EMI achieved by a finite-alphabet input in a rank-deficient channel yields a lower diversity order than the one achieved in a full-rank channel. This is similar to MIMO channels with Gaussian inputs under SST. Due to the page limitations, further discussions are skipped here and left as a potential direction for future work.

\section{Extension to Multi-Stream Transmission}\label{Section_MS}
For MST, the received signal vector is given by
{\setlength\abovedisplayskip{2pt}
\setlength\belowdisplayskip{2pt}
\begin{align}\label{MIMO_MST_Model}
{\mathbf{y}}=\sqrt{\bar\gamma}{\mathbf{H}}{\mathbf{P}}{\mathbf{x}}+{\mathbf{n}},
\end{align}
}where ${\mathbf{P}}\in{\mathbbmss{C}}^{N_{\rm{t}}\times N}$ denotes the precoding matrix satisfying ${\mathsf{tr}}\left\{{\mathbf{P}}{\mathbf{P}}^{\mathsf{H}}\right\}=1$ with $N$ being the number of data streams, and ${\mathbf{x}}\in{\mathbbmss{C}}^{N\times1}$ is the data vector with i.i.d. elements drawn from the $M$-ary constellation $\mathcal X$. Hence, the input signal $\mathbf{x}$ is taken from a multi-dimensional constellation ${\mathcal{Y}}$ consisting of $M^N$ points, i.e., $\mathbf{x}\in{\mathcal{Y}}=\left\{{\bm{\mathsf{x}}}_g\in{\mathbbmss{C}}^{N\times1}\right\}_{g=1}^{M^N}$, with $\mathbbmss{E}\left\{{\mathbf{x}}{\mathbf{x}}^{\mathsf{H}}\right\}={\mathbf{I}}_N$. Assume ${\bm{\mathsf{x}}}_g$ is sent with probability $q_g$, $0 < q_g < 1$, and the input distribution is given by ${\mathbf{q}}_{{\mathcal{Y}}}\triangleq[q_1,\cdots,q_{M^N}]\in{\mathbbmss{C}}^{1\times M^N}$ with $\sum_{g=1}^{M^N}q_g=1$. The MI in this case can be written as $\texttt{I}\left({\bar\gamma};{\mathbf{H}}{\mathbf{P}}\right)=H_{{\mathbf{q}}_{{\mathcal{Y}}}}-N_{\rm{r}}\log_2{\rm{e}}-
\sum\nolimits_{g=1}^{M^N}p_gf_{g}\left({\bar\gamma};{\mathbf{H}}{\mathbf{P}}\right)$, where $f_{g}\left({\bar\gamma};{\mathbf{H}}{\mathbf{P}}\right)\!\triangleq\!{\mathbbmss{E}}_{\mathbf{n}}\left\{
\log_2\left({\sum_{g'=1}^{M^N}\frac{p_{g'}}{p_g}{\rm{e}}^{-\left\|{\mathbf{n}}+\sqrt{\bar\gamma}{\mathbf{H}}{\mathbf{P}}{\mathbf{b}}_{g,g'}\right\|^2}}\right)
\right\}$ with ${\mathbf{b}}_{g,g'}={\bm{\mathsf{x}}}_g-{\bm{\mathsf{x}}}_{g'}=\left[b_{g,g',1},\cdots,b_{g,g',N}\right]^{\mathsf{H}}\in{\mathbbmss{C}}^{N\times1}$ \cite{Lozano2018}. Note that although the authors in \cite{Akemann2013} derived a closed-form expression for the MST-EMI achieved by Gaussian inputs, it is challenging to extend the results in \cite{Akemann2013} to systems with finite-alphabet inputs. We hence consider high-SNR limit while analyzing the MST-EMI achieved by finite-alphabet inputs.
\subsection{MST Without CSIT}
For the case without CSIT, the preocding matrix can be set to ${\mathbf{P}}=1/\sqrt{N_{\rm{t}}}{\mathbf{I}}_{N_{\rm{t}}}\triangleq{\mathbf{P}}_{\rm{no}}$ and the number of data streams is given by $N=N_{\rm{t}}$. The corresponding high-SNR MST-EMI is characterized in the following theorem.
\vspace{-5pt}
\begin{theorem}\label{EMI_Asy_MS_NCSIT_Asym_Basic}
Let $\bar\gamma\rightarrow\infty$. Then, the MST-EMI without CSIT can be characterized as ${\mathscr{I}}_{M}^{\mathcal{Y}}\simeq H_{{\mathbf{q}}_{{\mathcal{Y}}}}-{\mathcal{O}}\left({\bar\gamma}^{-1}\right)$.
\end{theorem}
\vspace{-5pt}
\begin{IEEEproof}
The proof is given in Appendix \ref{Proof_Theorem_EMI_Asy_MS_NCSIT_Asym_Basic}.
\end{IEEEproof}
\vspace{-5pt}
\begin{remark}
The results in Theorem \ref{EMI_Asy_MS_NCSIT_Asym_Basic} suggest that the diversity order of the MST-EMI without CSIT is given by ${\mathcal{G}}_{\rm{d}}=1$, which is the same as that of the SST-EMI without CSIT.
\end{remark}
\vspace{-5pt}
\subsection{MST With CSIT}
In this case, we consider two main precoding techniques.
\subsubsection{MRT Precoding}
By MRT precoding, we have
{\setlength\abovedisplayskip{2pt}
\setlength\belowdisplayskip{2pt}
\begin{align}
{\mathbf{P}}=\left[{\mathsf{tr}}\left({\mathbf{H}}^{\mathsf{H}}{\mathbf{H}}\right)\right]^{-1/2}{\mathbf{H}}^{\mathsf{H}}\triangleq {\mathbf{P}}_{\rm{mrt}}
\in{\mathbbmss{C}}^{N_{\rm{t}}\times N_{\rm{r}}}
\end{align}
}and $N=N_{\rm{r}}$, which yields ${\mathbf{y}}=\sqrt{\bar\gamma}{\mathbf{G}}{\mathbf{x}}+{\mathbf{n}}$ with ${\mathbf{G}}=\left\|{\mathbf{h}}_{\rm{t}}\right\|\frac{{\mathbf{h}}_{\rm{r}}}{\left\|{\mathbf{h}}_{\rm{r}}\right\|}{\mathbf{h}}_{\rm{r}}^{\mathsf{H}}
\in{\mathbbmss{C}}^{N_{\rm{r}}\times N_{\rm{r}}}$. The high-SNR EMI in this case is characterized as follows.
\vspace{-5pt}
\begin{theorem}\label{EMI_Asy_MS_CSIT_Asym_MRT}
Let $\bar\gamma\rightarrow\infty$. Then, the MST-EMI achieved by the MRT precoding satisfies ${\mathscr{I}}_{M}^{\mathcal{Y}}\simeq H_{{\mathbf{q}}_{{\mathcal{Y}}}}-{\mathcal{O}}\left({\bar\gamma}^{-1}\right)$.
\end{theorem}
\vspace{-5pt}
\begin{IEEEproof}
Similar to the proof of Theorem \ref{EMI_Asy_MS_NCSIT_Asym_Basic}.
\end{IEEEproof}
It is worth noting that the diversity order achieved by the MRT precoding is ${\mathcal{G}}_{\rm{d}}=1$, which is the same as that achieved without CSIT. To address this issue, we proceed to the max-$d_{\min}$ precoding scheme which enhances the diversity order.
\subsubsection{Max-$d_{\min}$ Precoding}
Optimizing $\texttt{I}\left({\bar\gamma};{\mathbf{H}}{\mathbf{P}}\right)$ at high-SNRs is equivalent to maximizing the
minimum distance \cite{Rodrigues2010}
{\setlength\abovedisplayskip{2pt}
\setlength\belowdisplayskip{2pt}
\begin{align}\label{d_min_def}
d_{\min}\triangleq\min\nolimits_{g\neq g'}\left\|{\mathbf{H}}{\mathbf{P}}{\mathbf{b}}_{g,g'}\right\|=\min\nolimits_{g\neq g'}\left|{\mathbf{h}}_{\rm{t}}^{\mathsf{H}}{\mathbf{P}}{\mathbf{b}}_{g,g'}\right|.
\end{align}
}The resulting max-$d_{\min}$ precoder is given by
{\setlength\abovedisplayskip{2pt}
\setlength\belowdisplayskip{2pt}
\begin{align}
\mathbf{P}_{\star}=\argmax\nolimits_{{\mathbf{P}}\in{\mathbbmss{C}}^{N_{\rm{t}}\times N},{\mathsf{tr}}\{{\mathbf{P}}{\mathbf{P}}^{\mathsf{H}}\}=1}d_{\min}.
\end{align}
}Yet, finding a closed-form solution to $\mathbf{P}_{\star}$ is a challenging task, which makes the subsequent analyses intractable. As a compromise, we propose a heuristic precoding design by exploiting the structure of $d_{\min}$. Specifically, by observing \eqref{d_min_def}, we design the heuristic max-$d_{\min}$ precoder as a rank-one matrix that satisfies ${\mathbf{P}}_{\rm{mm}}={\left\|{\mathbf{h}}_{\rm{t}}\right\|}^{-1}{\mathbf{h}}_{\rm{t}}{\mathbf{d}}_{\star}^{\mathsf{H}}\in{\mathbbmss{C}}^{N_{\rm{t}}\times N}$ with
{\setlength\abovedisplayskip{2pt}
\setlength\belowdisplayskip{2pt}
\begin{align}
{\mathbf{d}}_{\star}={\argmax}_{{\mathbf{x}}\in{{\mathbbmss{C}}^{N\times1}},\left\|{\mathbf{x}}\right\|=1}\min\nolimits_{g\neq g'}\left|{\mathbf{x}}^{\mathsf{H}}{\mathbf{b}}_{g,g'}\right|.
\end{align}
}Note that ${\mathbf{d}}_{\star}$ can be obtained via an off-line exhaustive search, since it is fixed in $\mathbf{H}$. The corresponding high-SNR MST-EMI is characterized in the following theorem.
\vspace{-5pt}
\begin{theorem}\label{EMI_Asy_MS_CSIT_Asym_Max_dmin}
Let $\bar\gamma\rightarrow\infty$. Then, the MST-EMI achieved by the heuristic max-$d_{\min}$ precoder can be characterized as ${\mathscr{I}}_{M}^{\mathcal{X}}\simeq H_{{\mathbf{q}}_{{\mathcal{Y}}}}-{\mathcal{O}}\left({\bar\gamma}^{-{\mathcal{G}}_{\rm{d}}}\right)$ with ${\mathcal{G}}_{\rm{d}}=\min\left\{N_{\rm{t}}m_{\rm{t}},N_{\rm{r}}m_{\rm{r}}\right\}$.
\end{theorem}
\vspace{-5pt}
\begin{IEEEproof}
Similar to the proof of Theorem \ref{EMI_Asy_MS_NCSIT_Asym_Basic}.
\end{IEEEproof}
\vspace{-5pt}
\begin{remark}\label{remark7}
In contrast to ${\mathbf{P}}_{{\rm{no}}}$ and ${\mathbf{P}}_{{\rm{mrt}}}$, the diversity order achieved by ${\mathbf{P}}_{{\rm{mm}}}$ can be improved by increasing $N_{\rm{r}}$ and $N_{\rm{t}}$, which highlights the superiority of the max-$d_{\min}$ precoder.
\end{remark}
\vspace{-5pt}
\subsection{Discussions on Keyhole Effect}
Consider the Rayleigh fading model. Since it is challenging to obtain a closed-form $\mathbf{P}$ that can maximize the MI with CSIT being available, we only consider the case without CSIT. Using the approach in deriving Theorem \ref{EMI_Asy_MS_NCSIT_Asym_Basic}, we find that the high-SNR MST-EMI without CSIT in full-rank and keyhole MIMO channels can be written as ${\mathscr{I}}_{M,{\rm{nk}}}^{\mathcal{X}}\simeq H_{{\mathbf{q}}_{{\mathcal{Y}}}}-{\mathcal{O}}\left({\bar\gamma}^{-N_{\rm{r}}}\right)$ and ${\mathscr{I}}_{M,{\rm{k}}}^{\mathcal{X}}\simeq H_{{\mathbf{q}}_{{\mathcal{Y}}}}-{\mathcal{O}}\left({\bar\gamma}^{-1}\right)$, respectively.
\vspace{-5pt}
\begin{remark}
Comparing ${\mathscr{I}}_{M,{\rm{nk}}}^{\mathcal{X}}$ with ${\mathscr{I}}_{M,{\rm{k}}}^{\mathcal{X}}$, we find the keyhole effect can reduce the diversity order of the MST-EMI. Similar conclusions can be given for other rank-deficient MIMO channels with finite-alphabet and Gaussian inputs under MST. We skip further details for sake of brevity.
\end{remark}
\vspace{-5pt}
\section{Numerical Results}
We now validate our analyses through numerical simulations. Here, we set $N_{\rm{t}}=N_{\rm{r}}=2$, $m_{\rm{t}}=2$, $m_{\rm{r}}=3$, $p_g = \frac{1}{M}$ for $g\in \{1,\cdots,M\}$, and $q_g = \frac{1}{M^N}$ for $g\in \{1,\cdots,M^N\}$. As a result, we have $H_{{\mathbf{p}}_{\mathcal{X}}}=\log_2{M}$ and $H_{{\mathbf{q}}_{\mathcal{Y}}}=N\log_2{M}$. The simulation results are gathered via $10^{6}$ channel realizations.

\subsubsection*{SST-EMI}
\begin{figure}[!t]
    \centering
    \subfigbottomskip=0pt
	\subfigcapskip=-5pt
\setlength{\abovecaptionskip}{0pt}
    \subfigure[Explicit results.]
    {
        \includegraphics[height=0.3\textwidth]{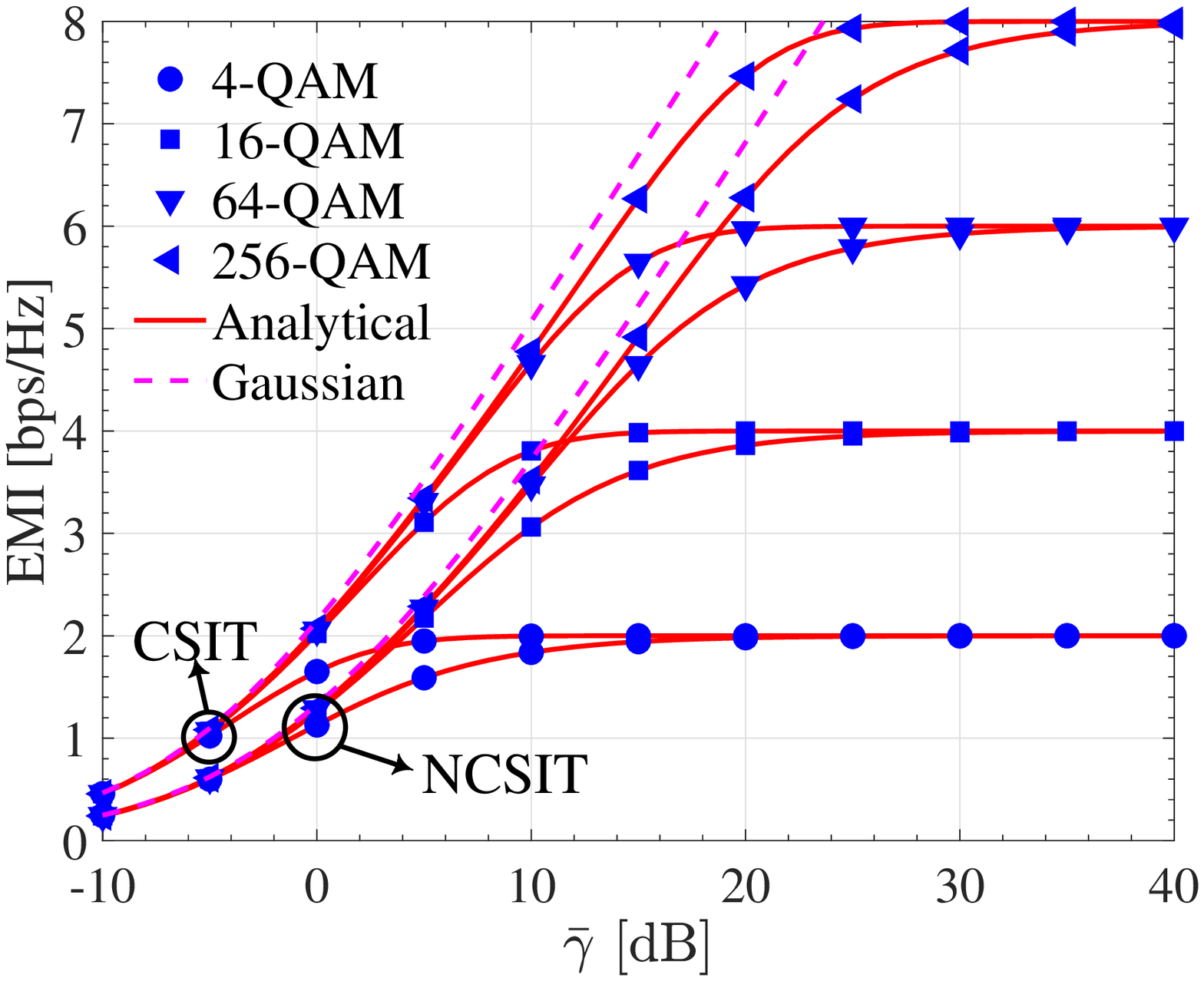}
	   \label{fig1a}	
    }
    \hspace{-13pt}
   \subfigure[Asymptotic results.]
    {
        \includegraphics[height=0.3\textwidth]{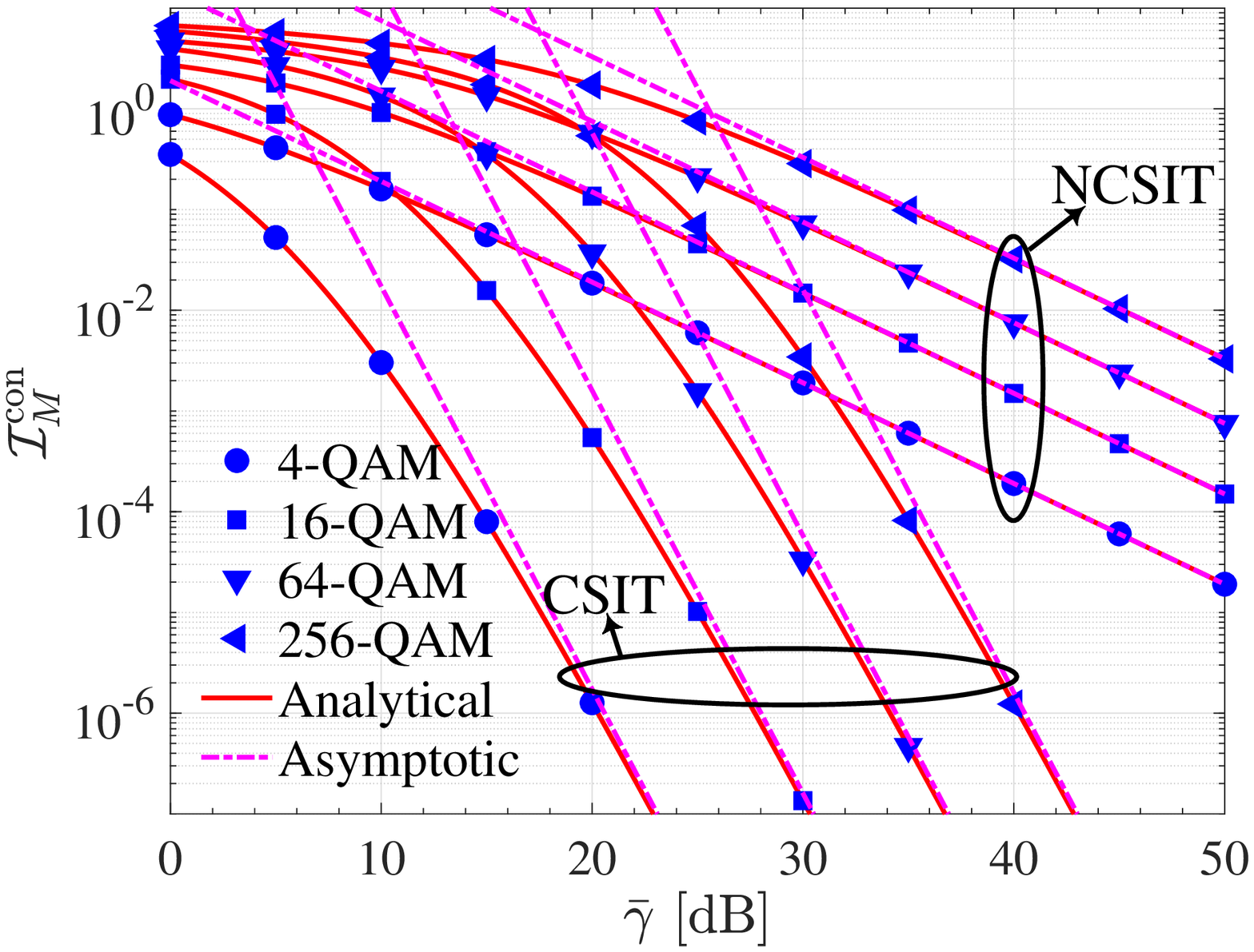}
	   \label{fig1b}	
    }
\caption{EMI of single-stream transmission.}
    \label{figure1}
    \vspace{-10pt}
\end{figure}

{\figurename} {\ref{fig1a}} shows the SST-EMI achieved by $M$-QAM signals for $M\in\{4, 16, 64, 256\}$ against the SNR, where the analytical EMI (denoted by solid lines) is calculated by \eqref{NCSIT_SS_EMI_Appr} or \eqref{CSIT_SS_EMI_Appr} by setting $V=200$. As {\figurename} {\ref{fig1a}} shows, the analytical results closely track the simulations (denoted by symbols). This verifies the accuracy of \eqref{NCSIT_SS_EMI_Appr} and \eqref{CSIT_SS_EMI_Appr}. For comparison, we also plot the EMI achieved by Gaussian signaling in {\figurename} {\ref{fig1a}}. As shown, the EMI of Gaussian inputs grows unboundedly as $\bar\gamma$ increases, whereas the EMI of finite-alphabet inputs converges to the entropy of the input, in the large limit of $\bar\gamma$. Moreover, we observe that the EMI with CSIT is higher than that without CSIT (denoted by NCSIT). By Remark \ref{remark1}, the rate of the EMI (${{{\mathcal I}}}_{M}^{\mathcal{X}}$) converging to $H_{{\mathbf{p}}_{\mathcal{X}}}$ equals the rate of ${\mathcal I}_{M}^{\rm {con}}=H_{{\mathbf{p}}_{\mathcal{X}}}-{{{\mathcal I}}}_{M}^{\mathcal{X}}$ converging to zero. To show this ROC, we plot ${\mathcal I}_{M}^{\rm {con}}$ versus $\bar\gamma$ in {\figurename} {\ref{fig1b}}. As shown, the derived asymptotic results almost perfectly match the numerical results in the high-SNR regime. This means that the diversity order derived in previous part is tight. It is further seen that the EMI with CSIT yields a faster ROC (or a higher diversity order) than that without CSIT. This agrees with the conclusion in Remark \ref{Compare_CSI}.

\begin{figure}[!t]
    \centering
    \subfigbottomskip=0pt
	\subfigcapskip=-5pt
\setlength{\abovecaptionskip}{0pt}
    \subfigure[Explicit results.]
    {
        \includegraphics[height=0.3\textwidth]{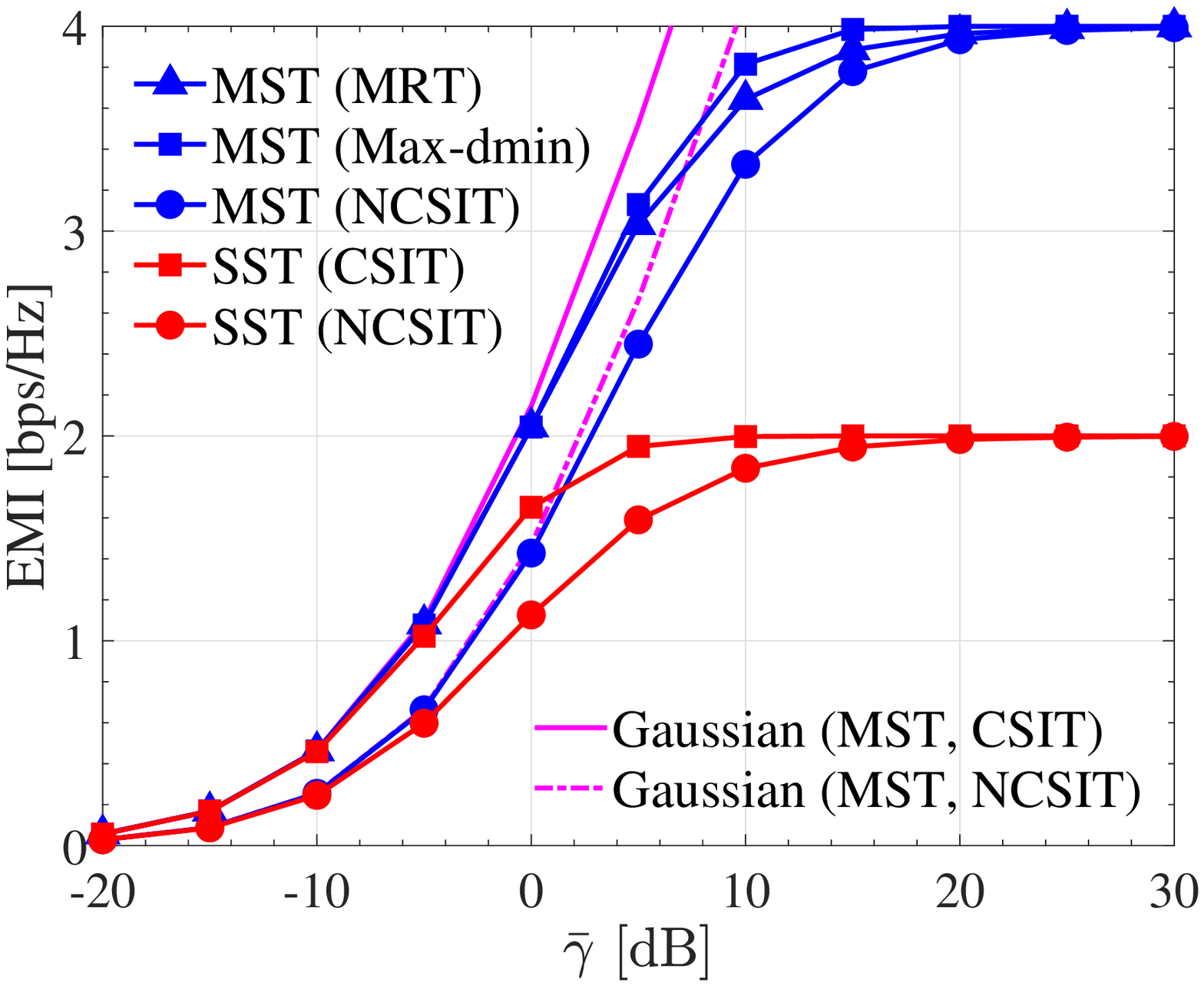}
	   \label{fig2a}	
    }
    \hspace{-13pt}
   \subfigure[Asymptotic results.]
    {
        \includegraphics[height=0.3\textwidth]{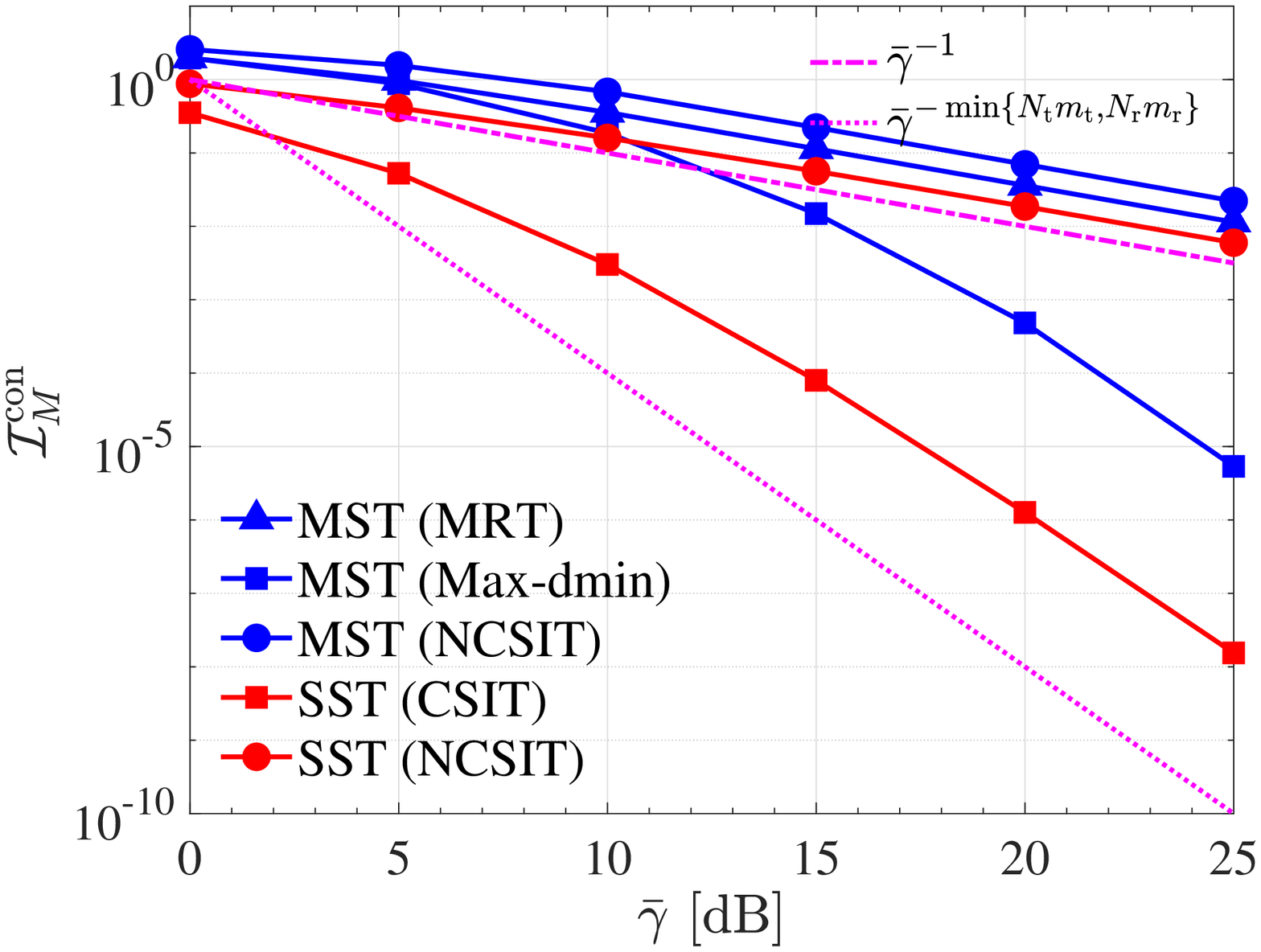}
	   \label{fig2b}	
    }
\caption{EMI of multi-stream transmission with $N=2$.}
    \label{figure2}
    \vspace{-10pt}
\end{figure}

\subsubsection*{MST-EMI}
Turn now to the MST-EMI. {\figurename} {\ref{fig2a}} compares the MST-EMI and SST-EMI achieved by 4-QAM and Gaussian signals. In both cases of with and without CSIT, the MST-EMI is higher than the SST-EMI and the Gaussian input achieves a higher EMI than finite-alphabet inputs. We further observe that the max-$d_{\min}$ precoding yields virtually the same EMI as the MRT precoder in the low-SNR regime but outperforms the latter one in the high-SNR regime. To show the ROC of the EMI, we plot ${\mathcal I}_{M}^{\rm {con}}=H_{{\mathbf{p}}_{\mathcal{X}}}-{{{\mathcal I}}}_{M}^{\mathcal{X}}$ (for SST) and ${\mathcal I}_{M}^{\rm {con}}=H_{{\mathbf{q}}_{\mathcal{Y}}}-{\mathscr{I}}_{M}^{\mathcal{X}}$ (for MST) versus $\bar\gamma$ in {\figurename} {\ref{fig2b}}. The curves for ${\bar\gamma}^{{-{\mathcal{G}}_{\rm{d}}}}$ are further provided to demonstrate the achievable diversity order. In the high-SNR regime, the curves for ${\mathcal I}_{M}^{\rm {con}}$ are parallel to ${\bar\gamma}^{{-{\mathcal{G}}_{\rm{d}}}}$. This indicates that the derived achievable diversity order is tight. Moreover, as {\figurename} {\ref{fig2b}} shows, the max-$d_{\min}$ precoder yields a faster ROC (or a higher diversity order) than the MRT precoder, which is consistent with the conclusion in Remark \ref{remark7}.

\section{Conclusion}
For keyhole MIMO channels with finite-alphabet inputs, irrespective of the number of streams, theoretical analyses indicate that the ROC of the EMI is determined by the array gain and the diversity order. It is further found that the keyhole effect can reduce the diversity order of the EMI achieved by finite-alphabet inputs.
\begin{appendices}
\section{Proof of Theorem \ref{Theorem_EMI_Asy_SS_NCSIT_Asym_Basic}}\label{Proof_Theorem_EMI_Asy_SS_NCSIT_Asym_Basic}
To facilitate the derivation, we rewrite the EMI as
{\setlength\abovedisplayskip{2pt}
\setlength\belowdisplayskip{2pt}
\begin{align}
{\mathcal{I}}_{M}^{\mathcal{X}}\!=\!\left.I_{M}^{\mathcal X}\left(t\right)F_{S_1}\left(\frac{N_{\rm{t}}t}{\bar\gamma}\right)\right|_{0}^{\infty}\!-\!\int_{0}^{\infty}\!\!F_{S_1}\left(\frac{N_{\rm{t}}t}{\bar\gamma}\right){\rm d}I_{M}^{\mathcal X}\left(t\right)
\end{align}
}with $F_{S_1}\left(\cdot\right)$ denoting the cumulative distribution function of $S_1$. According to \cite{Lozano2018}, we rewrite ${\mathcal{I}}_{M}^{\mathcal{X}}$ as
{\setlength\abovedisplayskip{2pt}
\setlength\belowdisplayskip{2pt}
\begin{align}
{\mathcal{I}}_{M}^{\mathcal{X}}\!
=\!H_{{\mathbf{p}}_{\mathcal{X}}}\!-\!\int_{0}^{\infty}\int_{0}^{{N_{\rm{t}}t}/{\bar\gamma}}f_{S_1}\left(x\right){\rm{d}}x\frac{{\rm{mmse}}_{M}^{\mathcal X}\left(t\right)}{\ln{2}}{\rm d}t,
\end{align}
}where ${\rm{mmse}}_{M}^{\mathcal X}\left(\gamma\right)=\frac{{\rm d}I_{M}^{\mathcal X}\left(\gamma\right)}{{\rm d}\gamma}\ln{2}$ is the MMSE in estimating $X$ in \eqref{AWGN_Channel} by observing $Y$ \cite{Lozano2018}. When $\bar\gamma\rightarrow\infty$, we have $\frac{1}{\bar\gamma}\rightarrow0$, which together with the facts of $K_{\nu}\left(z\right)\!=\!K_{-\nu}\left(z\right)$ \cite[Eq. (10.27.3)]{Paris2010} and $\lim_{z\rightarrow0}K_{\nu}\left(z\right)\!=\!\frac{1}{2}\Gamma\left(\nu\right)\left(\frac{1}{2}z\right)^{-\nu}$ ($\nu>0$) \cite[Eq. (10.30.2)]{Paris2010}, yields $\lim_{\bar\gamma\rightarrow\infty}{\mathcal{I}}_{M}^{\mathcal{X}}=\dot{\mathcal{I}}_{M}^{\mathcal{X}}$, where
{\setlength\abovedisplayskip{2pt}
\setlength\belowdisplayskip{2pt}
\begin{align}
\dot{\mathcal{I}}_{M}^{\mathcal{X}}\triangleq H_{{\mathbf{p}}_{\mathcal{X}}}\!-\!{\mathcal{F}}\!\left\langle\!{
\frac{\Gamma\left(\left|N_{\rm{r}}m_{\rm{r}}-h-1\right|\right)\hat{\mathcal{M}}\left(\bar{h}+1\right)}
{\Gamma\left(N_{\rm{r}}m_{\rm{r}}\right)\bar{h}\left({4U_{N_{\rm{t}}}\bar\gamma}/({N_{\rm{t}}m_{\rm{r}})}\right)^{\bar{h}}\ln{2}}\!}\right\rangle
\label{Asymptotic_EMI_SS_EMI}
\end{align}
}and $\bar{h}\triangleq\min\left\{N_{\rm{r}}m_{\rm{r}},h+1\right\}$. Then, we introduce the following two lemmas for further discussion.
\vspace{-10pt}
\begin{lemma}\label{Lemma_EMI_Asy_Aux1}
  Given the constellation ${\mathcal X}=\left\{\mathsf{x}_g\right\}_{g=1}^{M}$, the MMSE function satisfies $\lim_{\gamma\rightarrow\infty}{\rm{mmse}}_{M}^{\mathcal X}\left(\gamma\right)={\mathcal O}(\gamma^{-\frac{1}{2}}{\rm e}^{-\frac{\gamma}{8}d_{{\mathcal X},{\min}}^2})$, where $d_{\mathcal X,\min}\triangleq\min_{g\neq g'}\left|{\mathsf{x}_g}-{\mathsf{x}_{g'}}\right|$ \cite{Alvarado2014}.
\end{lemma}
\vspace{-10pt}
\vspace{-10pt}
\begin{lemma}\label{Lemma_EMI_Asy_Aux2}
  If $p\left(t\right)$ is ${\mathcal O}\left(t^a\right)$ as $t\rightarrow0^{+}$ and ${\mathcal O}\left(t^b\right)$ as $t\rightarrow+\infty$, then $\left|{\mathcal M}\left[p\left(t\right);z\right]\right|<\infty$ when $-a<z<-b$ \cite{Flajolet1995}.
\end{lemma}
\vspace{-10pt}
Particularly, $\lim_{t\rightarrow0^{+}}{\rm{mmse}}_{M}^{\mathcal X}\left(t\right)=1$ \cite{Lozano2018}, which together with Lemma \ref{Lemma_EMI_Asy_Aux1}, suggests that ${\rm{mmse}}_{M}^{\mathcal X}\left(t\right)$ is ${\mathcal O}\left(1\right)$ as $t\rightarrow0^{+}$ and ${\mathcal O}\left(t^{-\infty}\right)$ as $t\rightarrow\infty$. Using this fact and Lemma \ref{Lemma_EMI_Asy_Aux2}, we find that $|\hat{\mathcal{M}}\left(x\right)|<\infty$ holds for $0<x<\infty$, which in combination with the fact that ${\rm{mmse}}_{M}^{\mathcal X}\left(x\right)>0$ ($x>0$) \cite{Lozano2018}, suggests that $\hat{\mathcal{M}}\left(x\right)\in\left(0,\infty\right)$ holds for $0<x<\infty$. It follows from $\bar{h}=\min\left\{N_{\rm{r}}m_{\rm{r}},h+1\right\}>0$ that $\hat{\mathcal{M}}\left(\bar{h}+1\right)\in\left(0,\infty\right)$. As previously assumed, $N_{\rm{r}}m_{\rm{r}}\neq h+1$, $m_{\rm{r}}\geq\frac{1}{2}$, and $N_{\rm{r}}>1$, which yields $N_{\rm{r}}m_{\rm{r}}>1$. We then neglect the higher order terms in \eqref{Asymptotic_EMI_SS_EMI} to derive the asymptotic EMI as ${\mathcal{I}}_{M}^{\mathcal{X}}\simeq{H_{{\mathbf{p}}_{\mathcal{X}}}}-{\mathcal{G}}_{\rm{a}}^{-1}{\bar\gamma}^{-1}$, where ${\mathcal{G}}_{\rm{a}}^{-1}$ is shown in \eqref{Ga}.
\section{Proof of Theorem \ref{EMI_Asy_MS_NCSIT_Asym_Basic}}\label{Proof_Theorem_EMI_Asy_MS_NCSIT_Asym_Basic}
\begin{IEEEproof}
The MI satisfies \cite{Rodrigues2010}
\begin{align}
\texttt{I}\left({\bar\gamma};{\mathbf{H}}{\mathbf{P}}\right)=L\log_2{M}-\frac{1}{\ln{2}}\int_{{\bar\gamma}}^{\infty}{\text{mmse}}_{M}^{\mathcal X}\left(x;{\mathbf{H}}{\mathbf{P}}\right){\rm{d}}x,
\end{align}
where ${\text{mmse}}_{M}^{\mathcal X}\left(\bar\gamma;{\mathbf{H}}{\mathbf{P}}\right)$ denotes the MMSE in estimating $\mathbf{x}$ in \eqref{MIMO_MST_Model} by observing $\mathbf{y}$. Moreover, for any MIMO channels, the MMSE is bounded by \cite{Rodrigues2010}
\begin{align}
\underline{\text{mmse}}_{M}^{\mathcal X}\left(\bar\gamma;{\mathbf{H}}{\mathbf{P}}\right) \!\leq\!
{\text{mmse}}_{M}^{\mathcal X}\left(\bar\gamma;{\mathbf{H}}{\mathbf{P}}\right)           \!\leq\!
\overline{\text{mmse}}_{M}^{\mathcal X}\left(\bar\gamma;{\mathbf{H}}{\mathbf{P}}\right).
\end{align}
Defining $f_l\left(x\right)\triangleq1-\frac{1}{\sqrt{\pi}}\int_{-\infty}^{+\infty}
\tanh\left(\sqrt{x}a\right){\emph{e}}^{-{\left(a-\frac{\sqrt{x}}{2}\right)^2}}
{\rm{d}}a$, $f_u\left(x\right)\triangleq Q\left(\sqrt{\frac{{x}}{2}}\right)$ with $Q\left(x\right)\triangleq\frac{1}{\sqrt{2\pi}}\int_{x}^{\infty}{\emph{e}}^{-u^2/2}{\rm{d}}u$ being the Q-function, and $d_{i,k}\triangleq\left\|{\mathbf{H}}{\mathbf{P}}{\mathbf{b}}_{i,k}\right\|^2$, we have \cite[Appendix \uppercase\expandafter{\romannumeral3}]{Rodrigues2010}
\begin{align}
&\underline{\text{mmse}}_{M}^{\mathcal X}\left(\bar\gamma;{\mathbf{H}}{\mathbf{P}}\right)=\sum\nolimits_{i,k=1,k\neq i}^{M^L}
\frac{d_{i,k}}{4M^L}\frac{f_l\left(\bar\gamma d_{i,k}\right)}{M^L-1},\\
&\overline{\text{mmse}}_{M}^{\mathcal X}\left(\bar\gamma;{\mathbf{H}}{\mathbf{P}}\right)=\sum\nolimits_{i,k=1,k\neq i}^{M^L}\frac{d_{i,k}}{M^L}f_u\left(\bar\gamma d_{i,k}\right).
\end{align}
Therefore, the EMI is upper bounded by
\begin{align}
{\mathscr{I}}_{M}^{\mathcal{X}}\leq
L\log_2{M}-\frac{1}{\ln{2}}\int_{{\bar\gamma}}^{\infty}\underline{\text{mmse}}_{M}^{\mathcal X}\left(x;{\mathbf{H}}{\mathbf{P}}\right){\rm{d}}x\triangleq\overline{\mathscr{I}}_{M}^{\mathcal{X}}.
\end{align}
After some manipulations, we can get
\begin{align}
\overline{\mathscr{I}}_{M}^{\mathcal{X}}
=L\log_2{M}-\sum\nolimits_{i,k=1,k\neq i}^{M^L}\!\frac{\underline{\mathscr{I}}_{M,i,k}^{\mathcal{X}}\log_2{\emph{e}}}{4\left(M^L-1\right)M^L},
\end{align}
where $\underline{\mathscr{I}}_{M,i,k}^{\mathcal{X}}\triangleq\int_{0}^{\infty}\int_{{\bar\gamma}}^{\infty}{y}f_l\left(\bar\gamma y\right)f_{i,k}\left(y\right){\rm{d}}x{\rm{d}}y$ with $f_{i,k}\left(y\right)$ denoting the PDF of $d_{i,k}$. It follows that
\begin{align}
\underline{\mathscr{I}}_{M,i,k}^{\mathcal{X}}=\int_{0}^{\infty}\frac{1}{\bar\gamma}f_{i,k}\left(\frac{y}{\bar\gamma}\right)
\int_{y}^{\infty}f_l\left(x\right){\rm{d}}x{\rm{d}}y.
\end{align}
When ${\mathbf{P}}=1/\sqrt{N_{\text{t}}}{\mathbf{I}}_{N_{\text{t}}}$, we have $d_{i,k}=\left\|{\mathbf{h}}_{\text{r}}\right\|^2\left|1/\sqrt{N_{\text{t}}}{\mathbf{h}}_{\text{t}}^{\mathsf H}{\mathbf{b}}_{i,k}\right|^2$, whose PDF presents the same form as \eqref{NCSIT_SS_PDF} by setting $U_{N_{\text{t}}}=\sum_{a=1}^{N_{\text{t}}}\frac{\left|b_{i,k,a}\right|}{4m_{\text{t}}{N_{\text{t}}}}$ and $Y_{N_{\text{t}}}=\prod_{a=1}^{N_{\text{t}}}\left(\frac{\left|b_{i,k,a}\right|}{4m_{\text{t}}{N_{\text{t}}}}\right)^{i_a}$. Following similar steps as those outlined in Appendix \ref{Proof_Theorem_EMI_Asy_SS_NCSIT_Asym_Basic}, we find that when $\bar\gamma\rightarrow\infty$, it has
\begin{align}
\underline{\mathscr{I}}_{M,i,k}^{\mathcal{X}}\!\simeq\!\sum_{i_1=0}^{m_{\text{t}}-1}\!\!\cdots\!\!\sum_{i_{N_{\text{t}}}=0}^{m_{\text{t}}-1}
\!\frac{S_{N_{\text{t}}}!
Y_{N_{\text{t}}}m_{\text{r}}{\mathcal{M}}_l\left(1\right){\bar\gamma}^{-1}}
{4\left(N_{\text{r}}m_{\text{r}}-1\right)X_{N_{\text{t}}}U_{N_{\text{t}}}^{S_{N_{\text{t}}}+1}}
\end{align}
with ${\mathcal{M}}_l\left(t\right)\!\triangleq\!{\mathcal M}\!\left[\int_{y}^{\infty}\!f_l\left(x\right){\rm{d}}x;{t}\right]$. Define $\underline{f}\left(y\right)=\int_{y}^{\infty}f_l\left(x\right){\rm{d}}x$, and we have $\lim_{x\rightarrow0^{+}}f_l\left(x\right)=0$ and $\lim_{x\rightarrow\infty}f_l\left(x\right)={\mathcal{O}}\left({\emph{e}}^{-\frac{x}{4}}x^{-\frac{1}{2}}\right)$ \cite[Theorem 3, Appendix B]{Lozano2006}, indicating that $f_l\left(x\right)$ is ${\mathcal O}\left(x^{a}\right)$ ($a\geq0$) as $t\rightarrow0^{+}$ and ${\mathcal O}\left(x^{-\infty}\right)$ as $x\rightarrow\infty$. It follows form this fact and Lemma \ref{Lemma_EMI_Asy_Aux2} that $\lim_{y\rightarrow0^{+}}\underline{f}\left(y\right)=\int_{0}^{\infty}f_l\left(x\right){\rm{d}}x={\mathcal{O}}\left(y^0\right)\in\left(0,\infty\right)$. Moreover, based on L'H\^{o}spital's rule and \cite[Appendix B]{Lozano2006}, we can get $\lim_{y\rightarrow\infty}\underline{f}\left(y\right)={\mathcal{O}}\left({\emph{e}}^{-\frac{x}{4}}x^{-\frac{1}{2}}\right)$. By continuously using Lemma \ref{Lemma_EMI_Asy_Aux2}, we find that ${\mathcal{M}}_l\left(1\right)\in\left(0,\infty\right)$ and thus $\underline{\mathscr{I}}_{M,i,k}^{\mathcal{X}}={\mathcal{O}}\left({\bar\gamma}^{-1}\right)$. It follows that
\begin{align}\label{EMI_UB_Asym}
\lim_{\bar\gamma\rightarrow\infty}{\mathscr{I}}_{M}^{\mathcal{X}}\leq \lim_{\bar\gamma\rightarrow\infty}\overline{\mathscr{I}}_{M}^{\mathcal{X}}=N_{\text{t}}\log_2{M}-{\mathcal{O}}\left({\bar\gamma}^{-1}\right).
\end{align}
Turn now to the EMI's lower bound given by
\begin{align}
{\mathscr{I}}_{M}^{\mathcal{X}}\geq
L\log_2{M}-\sum\nolimits_{i,k=1,k\neq i}^{M^L}\!\frac{\overline{\mathscr{I}}_{M,i,k}^{\mathcal{X}}}{M^L\ln{2}}
\triangleq\underline{\mathscr{I}}_{M}^{\mathcal{X}}
\end{align}
with $\overline{\mathscr{I}}_{M,i,k}^{\mathcal{X}}\triangleq\int_{0}^{\infty}\int_{{\bar\gamma}}^{\infty}{y}f_u\left(\bar\gamma y\right)f_{i,k}\left(y\right){\rm{d}}x{\rm{d}}y$. We find that when $\bar\gamma\rightarrow\infty$,
\begin{align}\label{EMI_Lower_Bound_Basic}
\overline{\mathscr{I}}_{M,i,j}^{\mathcal{X}}\!\simeq\!\sum_{i_1=0}^{m_{\text{t}}-1}\!\!\cdots\!\!\sum_{i_{N_{\text{t}}}=0}^{m_{\text{t}}-1}
\!\frac{S_{N_{\text{t}}}!
Y_{N_{\text{t}}}m_{\text{r}}{\mathcal{M}}_u\left(1\right)U_{N_{\text{t}}}^{-S_{N_{\text{t}}}-1}{\bar\gamma}^{-1}}
{4\left(N_{\text{r}}m_{\text{r}}-1\right)\prod_{k=1}^{N_{\text{t}}}\left(\frac{\left(i_k!\right)^2}{\left(1-m_{\text{t}}\right)_{i_k}}\right)},
\end{align}
where ${\mathcal{M}}_u\left(t\right)\triangleq{\mathcal M}\left[\int_{y}^{\infty}f_u\left(x\right){\rm{d}}x;{t}\right]$. Then, following similar steps in proving ${\mathcal{M}}_l\left(1\right)\in\left(0,\infty\right)$, we can prove ${\mathcal{M}}_u\left(1\right)\in\left(0,\infty\right)$ and thus
\begin{align}
\lim_{\bar\gamma\rightarrow\infty}{\mathscr{I}}_{M}^{\mathcal{X}}\geq
\lim_{\bar\gamma\rightarrow\infty}\underline{\mathscr{I}}_{M}^{\mathcal{X}}=N_{\text{t}}\log_2{M}-{\mathcal{O}}\left({\bar\gamma}^{-1}\right),
\end{align}
which together with \eqref{EMI_UB_Asym}, yields $\lim_{\bar\gamma\rightarrow\infty}{\mathscr{I}}_{M}^{\mathcal{X}}=N_{\text{t}}\log_2{M}-{\mathcal{O}}\left({\bar\gamma}^{-1}\right)$.
\end{IEEEproof}
\end{appendices}

\end{document}